\documentstyle[aps,pra,twocolumn]{revtex}

\begin{document}
\author{M. I. Katsnelson\cite{mik}, V. V. Dobrovitski, and 
  B. N. Harmon}
\address{Ames Laboratory, Iowa State University, Ames, Iowa, 50011}
\title{Propagation of local decohering action in distributed quantum 
  systems}
\date{today}

\maketitle
\draft

\begin{abstract} 
We study propagation of the decohering influence caused by a local
measurement performed on a distributed quantum system. As an
example, the gas of bosons forming a Bose-Einstein condensate is
considered. We demonstrate that the local decohering perturbation
exerted on the measured region propagates over the system in the
form of a decoherence wave, whose dynamics is governed by
elementary excitations of the system. We argue that the 
post-measurement evolution of the system (determined by 
elementary excitations) is of importance for transfer of 
decoherence, while the initial collapse of the wave function 
has negligible impact on the regions which are not directly 
affected by the measurement.
\end{abstract}

\pacs{03.65.Bz, 05.30.Jp, 03.75.Fi}

\section*{Introduction} 
The theory of quantum measurement begun
in the 1920s still remains an active topic of interest (see, e.g.
Ref.\ \onlinecite{meas1} and references therein). According to
von Neumann's theory of measurement \cite{neumann}, unitary
evolution of a system prepared initially in a pure quantum state
is interrupted by an instant decohering action of the measuring
apparatus, so that the density matrix describing an ensemble of
such systems changes radically (it ceases to be a projection
operator) and entropy rises. This view has been shown to describe
rather accurately the consequences of an act of measurement, but
the dynamics of the measurement process itself is lacking.
The contemporary theory of quantum measurements, which provides much
deeper analysis of the measurement process, is based on the
concept of decoherence \cite{meas}. To be measured, the system
has to interact with its environment, which consists of a large
number of degrees of freedom. The Hilbert space of the system
becomes divided into subspaces corresponding to the same
eigenvalue of the system-environment interaction Hamiltonian. As
a result of this interaction, coherence between different
subspaces is quickly lost, and after the measurement the system
appears in a mixed state. The concept of decoherence turned out
to be successful in many areas of fundamental physics, such as
the study of macroscopic quantum effects \cite{leg} and
consistent histories interpretation of quantum mechanics
\cite{omnes}, so that investigation of this process and related
effects is of considerable importance.

At present, decoherence and its consequences for point-like
quantum systems have been studied in detail (for review, see 
Ref.\ \onlinecite{zurnew}), but distributed quantum systems 
have received
significantly less attention. Mostly, linear systems have been
investigated, where separation into noninteracting modes is
possible, and each mode is considered as an independent
oscillator \cite{zurfield}. However, this approach is difficult to 
apply to sufficiently
nonlinear systems (e.g., spin systems, or the Bose-Einstein
condensate as described by the Gross-Pitaevskii equation) possessing
localized soliton-like excitations. For systems where localized
excitations prevail, dealing explicitely with
real-space coordinates could be a more suitable strategy. 

A real-space description of decoherence in distributed systems is a
very general and complicated issue. In this paper, we consider
only one aspect of the problem, namely, how {\it local\/}
properties of different regions in a distributed quantum system
are affected by a {\it local\/} measurement, that acts only on
some part of the system. Indeed, different regions in the system
are not isolated from each other, and correlations between them
exist (or can build up). Therefore, in spite of the fact that a
local measurement initially affects only one region, other regions 
can ``acquire knowledge'' that some part of the system has been
measured. In this paper we explicitely show that the decohering
influence of the local measurement propagates through the system
in the form of a decoherence wave. Dynamics of the decoherence
wave is governed by elementary excitations, while the effect of
entanglement is very small for macroscopically large systems.

The consideration presented here can be applied to other similar
situations, so that a decoherence wave propagating with a
characteristic velocity of excitations is likely to be quite
common. This phenomenon, being a notable part of any real
measurement, is of fundamental interest. Moreover, propagation of
decoherence can be also of importance for the design of quantum
computers. Such a computer is a system of interacting quantum
entities, representing quantum bits (qubits). Fault-tolerant
quantum computations involve measurements performed on some
qubits and it is important to know how such measurement may
affect other qubits \cite{comput}. Moreover, decoherence is
introduced by a dissipative environment of qubits, so that
analysis of decoherence propagation may lead to strategies to
minimize influences detrimental to performance of the computer.

In this paper we consider a Bose-Einstein condensate of an ideal or
weakly non-ideal gas of bosons, which constitutes a good example
of a distributed system in a pure quantum state. It can be
implemented in reality as a gas of trapped atoms cooled down to
very low temperatures \cite{bose}. Suppose we measure the number
of particles in some region of space. If two such measurements are
done {\it simultaneously\/} at two different parts of the trap we
obtain the trivial result corresponding to the ground-state
wavefunction of the condensate. But if the second measurement is
carried out after some delay then the result is different and
provides information about the propagation of the perturbation
induced by the first measurement. 

The situation considered here is related to the problem of broken
gauge symmetry and existence of a relative phase of two
interfering condensates \cite{phase}, which has been extensively
discussed recently. If we have a condensate with a definite
number of particles, its phase is spread uniformly between 0 and
$2\pi$, while a definite phase requires a non-conservation of the
number of particles in the condensate. It has been shown that a
well-defined phase (evidenced experimentally by appearance of the
interference fringes) builds up in the course of the measurement
(atoms detection), due to increasing uncertainty in the number of
particles in each of the interfering condensates: each detected
atom may well belong to either of them. For the circumstances
considered in this paper, we have a similar situation: the local
phase of the condensate is the same in every region. Identity of
the phase throughout the condensate is due to uncertainty in the
local number of the particles inside each region. However, when
the number of particles in some region is determined by a local
measurement, the phase coherence in the condensate as a whole is
partially destroyed, what leads to observable consequences,
propagation of the decoherence wave in the system. Note that
decoherence wave is the same both for the condensate with
definite number of particles (with uncertain global phase, the
case of non-interacting bosons) and for the condensate with
definite global phase (but with uncertain number of condensed
particles, the case of weakly interacting bosons): the results
for the latter case transform exactly to the results for the
former as interaction goes to zero.

We describe the dynamics of the condensate in a linear approximation,
i.e. we use the approximation of noninteracting quasiparticles to
study a weakly non-ideal Bose-gas. In so doing, we loose the
ability to investigate some interesting nonlinear effects, but
we gain in clarity of presentation: it is reasonable to
start from a simplified (and not totally unrealistic) case to
emphasize the main idea.

We do not specify the way of measuring the local density of
condensate, and the dynamics of the measurement process is not
considered here. Analysis of a specific experimental scheme is
a distinct problem, requiring separate study, while here we 
focus on the post-measurement evolution of the condensate. In
principle, the local density of the Bose-condensate can be measured by
placing some probe into the trap, which interacts with the 
condensate so that an entangled state is formed
\begin{equation}
|X\rangle = \sum C_n |n\rangle \otimes |\alpha_n\rangle
\end{equation}
where $|n\rangle$ is the state of condensate with the number of
particles $n$ in the measured region, and $|\alpha_n\rangle$ is
the state of the probe. If the probe interacts with a large
number of environmental degrees of freedom, so that
$|\alpha_n\rangle$ are the eigenstates corresponding to different
eigenvalues of the probe-environment interaction Hamiltonian,
then the coherence between different probe states is being lost,
and the condensate's state also becomes an incoherent mixture of
different states $|n\rangle$. If the probe (and, correspondingly,
the condensate) decoheres quickly enough (as is usually the case) we
can consider the measurement as instantaneous and safely use von
Neumann's theory to describe the condensate's state immediately
after the measurement. 

Although the situation considered above is in many
respects too idealized to apply rigorously to a real experiment, it
is detailed enough to capture the essential processes of
concern in this paper. 

\section*{Propagation of decoherence in Bose-Einstein condensate}
To study quantitatively the effect of decoherence propagation,
let us consider first an ideal Bose-gas confined by external fields and
described by the Hamiltonian 
\begin{equation}
H = \sum_{\mu} E_{\mu} \alpha^{\dag}_{\mu} \alpha_{\mu},
\end{equation}
where $\alpha^{\dag}_{\mu}$ and $\alpha_{\mu}$ are the
boson creation and annihilation operators. $E_{\mu}$
are the one-particle energies, and we denote the
corresponding one-particle wavefunctions as 
$\varphi_{\mu}({\bf r})$, where $\mu =0$ stands for
the ground state having minimal energy $E_0=0$.
Then, the
ground-state eigenfunction of the system of $M$ bosons
can be written as
\begin{equation}
\label{ground}
|\Psi\rangle =\frac{1}{\sqrt{M!}} 
  \left(\alpha_{0}^{\dag}\right)^{M} |0\rangle,
\end{equation}
where $|0\rangle$ is the vacuum state. For simplicity, we can consider
the trap as being divided into a large number $N_c$ of small 
cells each having the
volume $V_0$ (it can be considered as the volume directly affected
by the measuring apparatus), satisfying the relation $V_0\ll V$,
where $V$ is the total volume of the trap. Then, the
coordinate ${\bf r}$ is understood as a discrete quantity 
(the number of a cell). This is similar to a general practice
in solid-state theory, where $V_0$ is analogous to the volume
of an elementary cell of the crystal \cite{ziman}. 
Note that in so doing, the
number of one-particle states taken into account becomes equal to
$N_c$, which is finite, though very large.
This corresponds to the fact that the number of states inside
the first Brillouin zone equals to the number of lattice cells.

At the instant $t=0$ we perform measurement
of the number of bosons in the cell ${\bf r}=0$. This observable
is represented by the operator $N=a^{\dag}(0) a(0)$,
where 
\begin{equation}
\label{sum}
a({\bf r}) = \sum\limits_{\mu} \varphi_{\mu}({\bf r}) \alpha_{\mu}.
\end{equation}
is the boson field operator. Eigenvalues of the operator $N$ 
are $n=0,1,2...$ and, suppose, the measurement
has given us one of them. According to von
Neumann's theory, it corresponds to the action
of the operator $W_n$ on the system, where
\begin{equation}
\label{w}
W_n=\delta _{n,N}=\int\limits_{0}^{2\pi} \frac{d\phi}{2\pi}
  \exp{\left[i\phi (n-N)\right]}
\end{equation}
is a projector onto the state with the number
of particles $n$ in the measured region. The operator $W_n$
has the value equal to unity on this state and it has
zero value on all others states.
Further development of the system is to be described by the density
matrix of the system $U(t)$, since the measurement interrupts unitary 
evolution and casts the system into mixed quantum state.
According to the standard theory of measurement \cite{neumann,meas},
the density matrix at the time $t$ is
\begin{equation}
\label{evolution} 
U(t)=\sum\limits_{n=0}^{\infty} \exp{(-iHt)} W_n U_{\text{in}} 
  W_n^{\dag} \exp{(iHt)} ,
\end{equation}
where $U_{\text{in}}=|\Psi\rangle \langle\Psi |$ 
is the density matrix before the measurement. 

To trace propagation of decoherence in the system, we 
study evolution of the one-particle density matrix 
\begin{equation}
\label{ro}
\rho({\bf r}, {\bf r}', t)=\mathop{\rm Tr}\left[ U(t) a^{\dag}({\bf r}') 
  a({\bf r}) \right]. 
\end{equation}
This quantity describes local properties of the Bose-Einstein condensate;
in particular, the average number of particles resulting 
from the second
measurement, which is performed at the point ${\bf r}$ at
the instant $t$, is given by the value $\rho({\bf r}, {\bf r}, t)$.

To simplify calculations, we use
the fact that the total number of particles is large,
$M\gg 1$, so that operators $\alpha_0$ and $\alpha^{\dag}_0$ acting
on the state $|\Psi\rangle$ can be replaced by the number $\sqrt{M}$
with relative accuracy $1/\sqrt{M}$; this is a standard approximation
in the theory of Bose-Einstein condensation \cite{agd}. Therefore,
Eq. (\ref{sum}) can be rewritten as 
\begin{equation}
\label{sum1}
a({\bf r})=\sqrt{n_B({\bf r})}+\bar a({\bf r}),
  \qquad \bar a({\bf r)} = \sum\limits_{\mu \neq 0}
  \varphi_{\mu}({\bf r}) \alpha_{\mu}
\end{equation}
where $n_B ({\bf r}) = M\varphi_0^2 ({\bf r})$ is the average number 
of condensate particles contained in the volume $V_0$ at the cell 
${\bf r}$. The expression for the one-particle density matrix 
can be written as
\begin{eqnarray}
\label{rho1}
\rho({\bf r}, {\bf r}', t) &=& \sum\limits_{n=0}^{\infty }
  \rho_n({\bf r},{\bf r}', t) ,\\
\nonumber
\rho_n({\bf r},{\bf r}', t) &=&
  \langle\Psi |W_n^{\dag} a^{\dag}({\bf r}',t)
  a({\bf r},t)W_n |\Psi\rangle,
\end{eqnarray}
where $a({\bf r}, t) = \exp{(iHt)} a({\bf r})\exp{(-iHt)}$. The
operator product in Eq.\ (\ref{rho1}) is to be ordered normally,
i.e. it is to be rewritten in such a way that 
all $a^{\dag}$ stand to the left of all $a$ in each term 
of the Taylor series expansion. In so doing, we take into account that
\begin{equation}
\label{commut}
\left[ a({\bf r}, t), \bar a^{\dag}(0)\right] = \sum\limits_{\mu\neq 0}
  \varphi_{\mu}({\bf r}) \varphi^*_{\mu}(0)
  {\rm e}^{-iE_{\mu}t} \equiv g({\bf r,}t).
\end{equation}
Note that for a system containing a large number of particles $M\gg 1$,
the function $g({\bf r}, t)$ can be replaced by the Green's function
\begin{equation}
G({\bf r},t) = \sum\limits_{\mu} \varphi_{\mu}({\bf r}) \varphi^*_{\mu}(0)
  {\rm e}^{-iE_{\mu}t}
\end{equation}
with accuracy of order of $1/M$, 
since $G({\bf r},t) = g({\bf r},t)+\varphi_0({\bf r})\varphi_0^*(0)$.
Performing the calculations, we obtain 
\begin{eqnarray}
\label{answer}
\rho_n({\bf r},{\bf r}', t) &=& p_n \left[\sqrt{n_B({\bf r})} 
  -  G({\bf r}, t) \sqrt{n_0} \right] \\
\nonumber
  &&\times \left[ \sqrt{n_B({\bf r}')}
  -  G^*({\bf r}', t) \sqrt{n_0} \right] \\
\nonumber
  &&+p_{n-1} n_0 G({\bf r}, t) G^*({\bf r}',t),
\end{eqnarray}
where $n_0=n_B(0)$, and $p_n = \mathop{\rm e}^{-n_0} n_0^n/(n!)$ 
is the Poisson distribution function.
Summation over $n$ can be performed 
explicitly, yielding
\begin{eqnarray}
\label{answ1}
\rho({\bf r},{\bf r}',t) &=& \sqrt{n_B({\bf r}) n_B({\bf r}')}
  -  G^*({\bf r}',t) \sqrt{n_B({\bf r}) n_0}\\
\nonumber
  && -  G({\bf r},t) \sqrt{n_B({\bf r}') n_0}
  + 2 n_0 G^*({\bf r}',t) G({\bf r},t).
\end{eqnarray}
This result shows that the measurement made at
the point ${\bf r=}0$
produces a decohering perturbation which propagates
over the trap in the form of a decoherence wave, and 
this propagation is
governed by the Green's function $G({\bf r},t)$. It
can be explicitely demonstrated by considering
an example of the gas 
consisting of free Bose-particles of mass $m$. The  
corresponding Green's function at the distances $r\gg V_0^{1/3}$
and times $t\gg mV_0^{2/3}/\hbar$ is \cite{feynman} 
\begin{equation}
\label{feyn1}
G({\bf r},t) = V_0 \left( \frac{m}{2\pi i\hbar t}\right)
  ^{3/2}\exp{\left( \frac{im{\bf r}^2}{2\pi \hbar t}\right)}.
\end{equation}
Local density of the
condensate after the measurement is given by the value
\begin{eqnarray}
\label{dens}
\rho({\bf r}, {\bf r}, t) &=& n_B + 2 n_B V_0^2 
  \left(\frac{m}{2\pi\hbar t}\right)^3 \\
 \nonumber
 && - 2 n_B V_0 \left( \frac{m}{2\pi\hbar t}\right)^{3/2} 
  \cos{\left(\frac{m {\bf r}^2}{2\pi\hbar t}\right)},
\end{eqnarray}
where $n_B=M/V$ is density of the condensate before the
measurement, which is independent on position $\bf r$.
This is an observable effect, which, in principle, can
be detected experimentally.

The entropy of the system, being initially zero, 
after the measurement is
\begin{equation}
S= -\mathop{\rm Tr} \left[ U(t)\ln{U(t)}\right] = 
  -\sum\limits_{n=0}^{\infty} p_n \ln{p_n} > 0,
\end{equation}
which is a clear indication of the decohering
effect of measurement. The increase of entropy of 
condensate as a whole happens only at the instant of measurement
and further evolution, being unitary, keeps it constant (decoherence
only propagates in the system from one region to another). Note that
local entropy (in contrast to the one-particle density matrix, where
the decoherence propagation is clearly seen) can be hardly used
to track the decoherence wave. The value of the local entropy 
is nonzero even in the initial pure state, while the total 
entropy of the system is zero. It happens because of ``negative 
entropy'' stored in the form of correlations between 
different parts of the condensate (for more detailed discussion
see Ref.\ \cite{zurinfo}).

The results obtained can be qualitatively interpreted as follows.
The measurement performed at ${\bf r}=0$ leads to localization
of some number of particles
within the cell ${\bf r}=0$. The localized particles
acquire rather large momenta, of order
$\hbar/V_0^{1/3}$; the average number of such particles is 
$n_0=n_B(0)$. Immediately after being localized, these
particles start to propagate over the trap, and their
propagation is governed by the Green's function (\ref{feyn1}).
Because of indistinguishability of particles in the trap,
we can not say that these are ``the same'' particles which
were measured at ${\bf r}=0$, so that the effect we consider
is not a physical motion of some separate particles in the trap,
but is the propagation of the decohering influence of
the measurement through the system. 

An interesting feature of the decoherence propagation
can be illustrated by the gas of bosons trapped
in a parabolic external potential, so that each particle is
represented by an isotropic harmonic oscillator of 
eigenfrequency $\Omega$. In this case, provided that 
$r\gg V_0^{1/3}$ and  $V_0\ll (\hbar /\Omega)^{3/2}\sim V$, 
the Green's function has the form \cite{feynman} 
\begin{equation}
\label{feynman}
G({\bf r},t) =V_0 \left(\frac{\Omega}{2\pi i\hbar 
  \sin{\Omega t}} \right)^{3/2} \exp{\left(\frac{i\Omega {\bf r}^2}
  {2\pi\hbar}\cot{\Omega t}\right)}
\end{equation}
where the particles are assumed to have unitary mass. This function
is periodic in time with the period $2\pi/\Omega$. 
Therefore, the decoherence propagation is also periodic 
in time with the same period. In the general case of Bose-gas
trapped in a finite volume, the decoherence propagation becomes
a quasiperiodic process, according to Eq.\ (\ref{commut}).

And, last but not least, decoherence propagation is a wave process,
possessing both amplitude and phase. Existence of coherent waves
in the system without quantum coherence is not
unusual, the same property is shared, e.g., by the sound wave
propagating in the classical fluid. Therefore, in principle, an
interference of two decoherence waves is possible.

Above, we have considered the system of noninteracting bosons.
Now, let us investigate the case of weakly interacting particles,
i.e. a weakly non-ideal Bose-gas contained in a trap of
large volume $V$. We assume no external potential acting on
the particles, so that the one-particle states 
are simple plane waves
\begin{equation}
\varphi_{\bf k}({\bf r}) = \sqrt{\frac{V_0}{V}} \exp{(i{\bf kr})},
\end{equation}
where the normalization reflects the fact that the
trap is divided into cells of volume $V_0\ll V$.
This system is described by the Hamiltonian 
\begin{eqnarray}
\label{nonideal}
H&=&\sum\limits_{\bf k} E_{\bf k} \alpha_{\bf k}^{\dag} \alpha_{\bf k}\\
\nonumber
  &&+ \frac 1{2V} \sum\limits_{{\bf k}_1+{\bf k}_2={\bf k}'_1+{\bf k}'_2}
  v ({\bf k}_1-{\bf k}'_1) \alpha_{{\bf k}'_1}^{\dag} 
  \alpha_{{\bf k}'_2}^{\dag} \alpha_{{\bf k}_2} \alpha_{{\bf k}_1}
\end{eqnarray}
where $v({\bf k})$ is the Fourier transform of the interaction
potential (which is assumed to be repulsive). Since the interaction
is small, new Bose operators can be introduced according to 
Bogoliubov transformation 
\begin{eqnarray}
\label{bogol}
\alpha_{\bf k} &=& \xi_{\bf k} \cosh{\chi_{\bf k}} + 
  \xi_{-{\bf k}}^{\dag} \sinh{\chi_{\bf k}}, \\
\nonumber
\alpha_{-{\bf k}}^{\dag} &=& \xi_{\bf k} \sinh{\chi_{\bf k}} + 
  \xi_{-{\bf k}}^{\dag} \cosh{\chi_{\bf k}},
\end{eqnarray}
with the parameters $\chi_{\bf k}$ defined as 
\begin{equation}
\label{bogol1}
\tanh{2\chi_{\bf k}} = -\frac{v({\bf k})n_B}{E_{\bf k} + 
  v({\bf k})n_B},
\end{equation}
where $n_B$ is the average number of particles belonging to Bose-Einstein
condensate contained in the volume $V_0$. Provided that the interaction
is small (or the gas density $M/V$ is small), almost all
particles belong to the condensate, so we can take
$n_B= M V_0/V$ with relative accuracy of order of $\sqrt{v(0) M/V}$
\cite{agd}.
By using the Bogoliubov transformation, we pass to the ideal gas
of new excitations with the dispersion law 
\begin{equation}
\label{bogol2}
\omega_{\bf k}=\sqrt{E_{\bf k}^2 + 2 E_{\bf k} v({\bf k}) n_B}.
\end{equation}

Again, we consider dynamical behavior of the one-particle density matrix.
The calculation procedure remains essentially the same as for 
the ideal Bose-gas. In so doing, we obtain the result:
\begin{eqnarray}
\label{answnew}
\rho_n ({\bf r},{\bf r}',t) &=& \frac{n_B}{(n!)^2}
  \frac{\partial^{2n}}{\partial z^n \partial z^{\prime n}}
  \Bigl\{ [1+(z-1) G({\bf r},t)] \\ 
\nonumber
  &&\times [1+(z'-1) G^*({\bf r}',t)] \\
\nonumber
  &&\times\exp{[n_B X(z,z')]}\Bigr\}_{z=z'=0}
\end{eqnarray}
where the following notations were used,
\begin{eqnarray}
\label{answlast}
X(z,z') &=& B(zz'-1) + (1-B)(z+z'-2) \\
  && + A\left[ (z-1)^2+(z'-1)^2\right], \\
\nonumber
A &=& \frac{V_0}{2V} \sum\limits_{\bf k} \frac{v({\bf k})n_B}
  {\omega_{\bf k}}, \\
\nonumber
B &=& \frac{V_0}{2V} \sum\limits_{\bf k} \left[1+\frac{E_{\bf k}
  +v({\bf k}) n_B} {\omega_{\bf k}}\right],
\end{eqnarray}
and $G({\bf r},t)$ is the Green's function of the weakly interacting
Bose-gas:
\begin{eqnarray}
\label{green}
G({\bf r},t) &=& \sum\limits_{\bf k} \exp{(i{\bf kr})}\\
\nonumber
  &&\times \left\{ \cos{\omega_{\bf k}t}
  -i\frac{E_{\bf k} + v({\bf k}) n_B}{\omega_{\bf k}}
  \sin{\omega_{\bf k}t} \right\}.
\end{eqnarray}
Again, we see that the decoherence wave propagating in the
system follows the dynamics of the 
Green's function (\ref{green}). Dynamic behavior of $G({\bf r},t)$ 
at large times $t$ and large distances $r$ 
can be analyzed by the
method of stationary phase \cite{witham}. According to
this method, the value
of the function $G(r,t)$ at the point $\bf r$ at the
instant $t$ is determined by those excitations which have 
a group velocity ${\bf u}({\bf k})\equiv d\omega_{\bf k}/ d{\bf k}$
obeying the requirement ${\bf u}({\bf k}) = {\bf r}/t$.
The excitations with large wavevectors ${\bf k}$ are 
subject to considerable damping
\cite{agd}, so that at large distances only the undamped long-wavelength
excitations determine the dynamics of the Green's function.
These excitations represent
sound propagating in the Bose-gas with the velocity $c=\sqrt{n_B v(0)/m}$,
so the decoherence wave 
in a system of weakly interacting bosons propagates with the 
sound velocity $c$.

This result can be interpreted in the same way as the 
decoherence wave in an ideal Bose-gas. The measurement
affects the particles situated at ${\bf r}=0$. Due to 
the interparticle interaction, the decohering perturbation 
is transferred to other regions of the system.
The decoherence transfer is governed by the 
undamped excitations present in the system, i.e. by the 
long-wavelength excitations traveling with
the sound velocity $c$.

\section*{Discussion} 
Summarizing, we have studied the decohering
influence of a local measurement performed on a distributed
quantum system. We show that the decohering perturbation exerted
on the measured region propagates over the system by forming a
{\it decoherence wave\/}, whose dynamics is determined by the
Green's function of the system. This result, although 
not totally unexpected, is not as trivial as it might seem, since
decoherence is a rather peculiar effect, and the decohering impact
of a measurement can be quite different from other physical
influences (see, e.g. the discussion in Ref.\ \cite{scully}). 

The usual scenario for few-particle systems is
based on the Einstein-Podolsky-Rosen (EPR) situation \cite{epr}
of strong entanglement, when, e.g. two
particles with spins $1/2$ form a singlet state
\begin{equation}
|\psi\rangle = \frac 1{\sqrt{2}}\left(\,|\!\uparrow\downarrow\rangle 
  - |\!\downarrow\uparrow\rangle\, \right).
\end{equation}
If the first spin has been measured, and as a result of this
measurement has been cast in the state $|\!\uparrow\rangle$ (here
again we use von Neumann's theory of instant measurement), then
the transfer of decoherence is instant: the second spin
immediately occurs in the state $|\!\downarrow\rangle$. In
distributed systems this effect is also present: the wave function of
the system collapses immediately after the measurement. But the
impact of the collapse upon the one-particle density matrix 
(and even $s$-particle density matrix, for $s\ll M$) is
practically unobservable for the system of macroscopic size
(where $M\gg 1$): the change in the density matrix element
$\rho({\bf r},{\bf r}',t)$ immediately after the measurement is
of order of $n_0/M$ (provided, of course, that ${\bf r},{\bf r}
'\neq 0$), and the same is true for the $k$-particle density
matrix 
$\rho({\bf r}_1,\dots{\bf r}_k; {\bf r}'_1,\dots{\bf r}'_k)$ 
if $k\ll M$. This result is rather obvious: localization of the
number $n_0$ of particles in some cell can not affect noticeably
other cells if the total number of particles is macroscopically
large. Therefore, the post-measurement evolution of the system,
which is governed by the Green's function, becomes important since it
provides much more noticeable changes in the density
matrix elements: in Eq.\ (\ref{dens}) the term corresponding to
the decoherence wave does not go to zero as $M\to\infty$.
Obviously, it happens because in the EPR-like situation the
entanglement is very ``stiff'', so that each state of one
particle determines completely the state of the other. But in the
many-particle system there is no one-to-one correspondence,
since the total number of degrees of freedom is much larger than
the number of degrees of freedom fixed during the measurement.
This difference is the reason for the different dynamics of
decoherence propagation.

Finally, we remark that another aspect of decoherence in
distributed systems has been studied within the context of
decoherent quantum histories \cite{gellmann,brun}. Although the
effects studied there, as well as systems considered and methods
used, are different from those investigated here, it is
interesting to note that local properties of distributed quantum
systems are often ``intrinsically'' decoherent \cite{brun} if a
coarse enough description is used. For the effects considered
here, sufficient coarse graining leads to averaging of the
oscillating Green's function over the spatial scale of several
oscillations, so that the details of the decoherence wave becomes
negligible. Therefore, the intrinsic structure of the decoherence
wave can be distinguished only at fine scales, where coherence of
the Green's function holds.

This work was partially carried out at the Ames Laboratory, which 
is operated for the U.\ S.\ Department of Energy by Iowa State 
University under Contract No.\ W-7405-82 and was supported by 
the Director for Energy Research, Office of Basic Energy Sciences 
of the U.\ S.\ Department of Energy.

\end{document}